\documentclass[epj,referee]{svjour}
\usepackage{bm,amsmath,amssymb,latexsym,mathrsfs,enumerate}
\usepackage[mathcal]{euscript}
\usepackage{hyperref}
\usepackage{graphicx}
\usepackage[numbers,sort&compress]{natbib}
\renewcommand{\vec}[1]{\bm{#1}}
\begin{document}

\title{Dynamics of topological solitons in two--dimensional ferromagnets}

\titlerunning{Dynamics of topological solitons in two--dimensional ferromagnets}

\author{Denis~D.~Sheka\inst{1}, Christian~Schuster\inst{2},  Boris~A.~Ivanov\inst{3},
and Franz~G.~Mertens\inst{2}}

\authorrunning{Sheka, Schuster, Ivanov, Mertens}

\offprints{Denis~D.~Sheka}

\mail{Denis~D.~Sheka}

\institute{National Taras Shevchenko University of Kiev, 03127 Kiev,
Ukraine, \email{Denis\_Sheka@univ.kiev.ua} \and Physics Institute,
University of Bayreuth, 95440 Bayreuth, Germany, \and Institute of
Magnetism, 04071 Kiev, Ukraine}

\date{Received: \today}

\abstract{ Dynamical topological solitons are studied in classical
two--dimensional Heisenberg easy--axis ferromagnets. The properties
of such solitons are treated both analytically in the continuum
limit and numerically by spin dynamics simulations of the discrete
system. Excitation of internal mode causes orbital motion. This is
confirmed by simulations.
 % end of PACS codes
} %end of abstract
\maketitle

\section{Introduction}
\label{sec:introduction}

The analysis of two--dimensional (2D) magnetic solitons continues
for more than 25 years, for reviews see
Refs.~\cite{Kosevich90,Bar'yakhtar94,Mertens00}. Such solitons are
well--known to play an important role in the physics of 2D magnetic
systems. In easy--plane magnets with continuously degenerated ground
state there appear magnetic vortices, which are responsible for the
Berezinski\u\i--Kosterlitz--Thouless phase transition
\cite{Berezinsky72,Kosterlitz73}. Be\-lavin and Polyakov were the
first who constructed exact analytical solutions for 2D topological
solitons in the \emph{isotropic} magnet in the continuum limit, and
proved that such solitons are responsible for the destruction of the
long--range ordering for finite temperature \cite{Belavin75}. In the
\emph{ani\-sotropic} magnets such static solitons are unstable
against collapse \cite{Hobard63,Derrick64}. However in easy--axis
magnets there appear various types of dynamical localized
topological solitons due to the presence of additional integrals of
motion. We will consider \emph{precessional} solitons
\cite{Kosevich90}, which exist in uniaxial magnets due to the
conservation of the $z$--projection ${S}_z$ of the total spin
\cite{Kosevich90,Bar'yakhtar93}. Precessional solitons  are known
for a number of models used in field theory and condensed matter
physics, see Ref.~\cite{Makhankov78}. The topological small radius
solitons become interesting now due to possible applications in
high--energy physics \cite{Walliser00} and the quantum Hall effect
\cite{Prange90}.

The problem of the dynamics of topological solitons and vortices is
a complicated task for 2D ferromagnets, where Lorentz and Galilean
invariance are absent. The presence of a gyroforce acting on a
moving soliton is the only thing which is well established, but the
free gyroscopic dynamics has not been reported till now for any 2D
solitons. For the easy--plane magnets the weak localization of the
vortex is related to the gapless magnon spectrum; hence the vortex
dynamics is governed mostly by the interaction with the system
border, and the inertial properties do not appear. As a result,
computer simulations of magnetic vortex dynamics in a large but
finite lattice show a motion which can be described by complicated
non--Newtonian dynamical equations with nonlocal terms
\cite{Ivanov98}. For the isotropic magnet with the gapless magnon
dispersion law numerical analysis shows the absence of the localized
motion of the soliton \cite{Papanicolaou95}.

A very attractive candidate to discuss the general problems of the
magnetic soliton motion is the easy--axis ferromagnet. In this case
the soliton shape is exponentially localized, it seems to be
possible to separate the soliton motion from the magnons due to
their finite activation energy. The general features of the 2D
soliton dynamics, which should have particle--like properties with a
finite soliton mass, are not clear at present. For example, in works
of Papanicolaou \emph{et al}
\cite{Papanicolaou91,Papanicolaou95,Papanicolaou96} the dynamics of
2D solitons was described using the algebra for some noncanonic
momentum. In particular, as it was mentioned in
Ref.~\cite{Papanicolaou91}, a single topological soliton can not
move without external field. At the same time a free rotational
motion of the 2D topological soliton was predicted in
Ref.~\cite{Ivanov89}. It results in a finite mass for the
small--radius soliton \cite{Ivanov89}, while the mass of the
localized soliton diverges as the logarithm of the system size
according to \cite{Zaspel93}.

The present work is devoted to the analysis of the dynamical
properties of topological solitons in easy--axis ferromagnets, both
in the discrete model with weak anisotropy and in the continuum
model. We should stress here that topological solitons were studied
only in the frameworks of continuum field approaches in all above
mentioned papers. Spin dynamics simulations for the motionless 2D
topological solitons were performed in \cite{Kamppeter01}. In this
paper we perform spin dynamics simulations for a wide range of
soliton shapes: from large radius solitons to small ones. The main
issue is to move the soliton. Using the structure of internal modes
\cite{Sheka01}, we have found such perturbations of initial
centrosymmetric soliton shape, which results in perfect orbital
motion of the soliton.

\section{Discrete Model and Continuum Limit for 2D Ferromagnets}
\label{sec:model}

We consider the simplest model of the classical 2D ferromagnet
described by the following Hamiltonian
\begin{equation} \label{eq:H-discrete}
\mathscr{H} = -\frac{J}{2}\!\!
\sum_{\left(\vec{n},\vec{\alpha}\right)}\!
\left(\vec{S}_{\vec{n}}\!\cdot\! \vec{S}_{\vec{n}+\vec{\alpha}} +
\delta S^z_{\vec{n}} S^z_{\vec{n}+\vec{\alpha}} \right).
\end{equation}
Here  $\vec{S}_{\vec{n}}\equiv\left(S^x_{\vec{n}}, S^y_{\vec{n}},
S^z_{\vec{n}}\right)$ is a classical spin vector with fixed length
$S$ (in units of the Plank constant $\hslash$) on the site $\vec{n}$
of a two--dimensional square lattice, $\vec{\alpha}$ is a vector to
a nearest neighbor. The model includes the  isotropic Heisenberg
exchange interaction, $J
> 0$ is the exchange integral, and the spatially  homogeneous uniaxial exchange
anisotropy, $\delta$ is the anisotropy constant.  The summation runs
over nearest--neighbor pairs $(\vec{n},\vec{n}+\vec{\alpha})$. The
case $\delta=0$ corresponds to the isotropic model. To describe the
anisotropy effects we will consider the case when $\delta>0$, then
the $z$--axis suppose the easiest magnetization.

The spin dynamics is described by the discrete version of the
Landau--Lifshitz equations
\begin{equation} \label{eq:LL-discrete}
\frac{d \vec{S}_{\vec{n}} }{dt} =  - \frac1\hslash \left[
\vec{S}_{\vec{n}}\times \frac{\partial \mathscr{H} }{\partial
\vec{S}_{\vec{n}}}\right].
\end{equation}
The model of the pure uniaxial ferromagnet has well--known linear
excitations (magnons) above the ground state $\vec{S}^z_{\vec{n}}=1$
of the form $1-\vec{S}^z_{\vec{n}}=\text{const}\ll1$,
$\vec{S}^x_{\vec{n}}+i\vec{S}^y_{\vec{n}}\propto
\exp(ik_xa+ik_ya-i\omega t)$, which have the finite gap dispersion
law
\begin{equation}\label{eq:magnons_discrete}
\omega(\vec{k}) = \omega_0 + \frac{4J S}{\hslash}\left[\sin^2
\left(\frac{k_xa}{2}\right)+\sin^2
\left(\frac{k_ya}{2}\right)\right].
\end{equation}
Here $\omega_0 = 4JS\delta/\hslash$ is the homogeneous ferromagnetic
resonance frequency, $\vec{k}$ is the wave vector.

In the case of weak anisotropy, $\delta \ll  1$, the characteristic
size $l_0 = a/ \sqrt{4\delta}$ of the excitations is larger than the
lattice constant $a$, so that in the lowest approximation in the
small parameter $a/l_0$ and with weak gradients of magnetization one
can use the continuum approximation for the Hamiltonian
\eqref{eq:H-discrete} by introducing the normalized spin $\vec{s} =
{\vec{S}}/{S} =
\left(\sin\theta\cos\phi;\sin\theta\sin\phi;\cos\theta\right).$ The
continuum version of the Hamiltonian is
\begin{equation}\label{eq:Energy}
\mathscr{E}[\theta,\phi] = \frac{JS^2}{2}\!\!\int\!\! \mathrm{d}^2 x
\Biggl[\left(\nabla\theta\right)^2 +
\sin^2\theta\left(\nabla\phi\right)^2 +
\frac{\sin^2\theta}{l_0^2}\Biggr].
\end{equation}
In terms of the fields $\theta$ and $\phi$, the Landau--Lifshitz
equations \eqref{eq:LL-discrete} read
\begin{equation} \label{eq:LL-continuum}
\sin\theta\ \partial_t \phi = - \frac{a^2}{\hslash S} \frac{\delta
\mathscr{E}}{\delta \theta}, \qquad \sin\theta\ \partial_t \theta =
\frac{a^2}{\hslash S}\frac{\delta \mathscr{E}}{\delta \phi}.
\end{equation}
In the longwavelength limit the magnon excitations of the form
$\theta = \text{const}\ll1, \quad \phi = \vec{k}\cdot \vec{r} -
\omega t$, have the following dispersion law,
\begin{equation}\label{eq:magnons}
\omega(\vec{k}) = \omega_0(1+ k^2l_0^2),
\end{equation}
which follows from \eqref{eq:magnons_discrete} in the lowest
approximation in $ka\ll1$.

\section{The Structure of Precessional Soliton}
\label{sec:structure}

For the pure uniaxial ferromagnet  the Hamiltonian
\eqref{eq:H-discrete} does not depend explicitly on the variable
$\phi$ due to the spin--space isotropy (in contrast to the lattice,
which is always anisotropic but in coordinate space). This condition
corresponds to the additional integral of motion
\begin{equation} \label{eq:N-discrete}
N = \sum_{\vec{n}} \left(S-S^z_{\vec{n}} \right).
\end{equation}
When $N\gg 1$ and the WKB approach is valid, one can consider
$N\in\mathbb{N}$ as the number of magnons,  bound in the soliton,
see Ref.~\cite{Kosevich90}. The conservation law
\eqref{eq:N-discrete} can provide a conditional minimum of the
Hamiltonian, which stabilize the possible soliton solution, see
below. The continuous version of \eqref{eq:N-discrete} reads
\begin{equation}\label{eq:N}
N = \frac{S}{a^2}\int \mathrm{d}^2 x \left(1-\cos\theta\right).
\end{equation}

The simplest nonlinear excitation of the  model
\eqref{eq:LL-continuum} is a 2D soliton, which has a finite energy.
The topological properties of the soliton are determined by the
mapping of the  $xy$--plane to the $S^2$--sphere of the order
parameter space. This mapping is described by the homotopic group
$\pi_2(S^2) = \mathbb{Z}$, which is characterized by the topological
invariant (Pontryagin index)
\begin{equation} \label{eq:Pontryagin}
\begin{split}
q &= \frac 1{4\pi }\int \mathrm{d}^2 x \mathcal{Q}, \qquad
\mathcal{Q} = \frac{\varepsilon _{\alpha \beta }}{2}
\Bigl[\vec{s}\cdot\bigl(\nabla _\alpha \vec{s}\times \nabla _\beta
\vec{s}\bigr)\Bigr].
\end{split}
\end{equation}
The Pontryagin index takes integer values, $q\in\mathbb{Z}$, being
an integral of motion.

Let us consider the so--called \emph{centrosymmetric topological
precessional soliton}, which has the following structure:
\begin{equation} \label{eq:cs-soliton}
\theta = \theta_0\left(\rho\right), \qquad \phi = \varphi_0 + q\chi
- \omega_p t,
\end{equation}
where $\rho=r/l_0$ is the dimensionless radius and
$\omega_p\in(0,\omega_0)$ is the frequency of the internal
precession. We will discuss only the case $q=1$, when the soliton
has a lower energy. The form of the function $\theta_0(\bullet)$
satisfies the following differential problem:
\begin{subequations} \label{eq:m-static}
\begin{align} \label{eq:m-static(1)}
&\frac{\mathrm{d}^2\theta_0}{\mathrm{d}{\rho}^2}+\frac1\rho
\frac{\mathrm{d}\theta_0}{\mathrm{d}{\rho}} - \sin\theta_0\cos\theta
\left(1+\frac{1}{{\rho}^2}\right)
+ \frac{\omega_p}{\omega_0} \sin\theta_0=0, \\
\label{eq:m-static(2)} %
&\theta_0(0)=\pi, \qquad \theta_0(\infty)=0.
\end{align}
\end{subequations}
This equation was solved numerically in
Refs.~\cite{Kovalev79,Kosevich81,Voronov83}. For the case of a
centrosymmetric soliton the number of bound magnons
\begin{equation}\label{eq:N'}
N = N_0 \int_0^\infty\!\!\! \rho\,\mathrm{d}\rho
\left[1-\cos\theta_0(\rho)\right],
\end{equation}
where $N_0=2\pi Sl_0^2/a^2$ is the characteristic number of bound
magnons for 2D magnets \cite{Kovalev79}. Multiplying
Eq.~\eqref{eq:m-static(1)} by
$\rho^2{\mathrm{d}\theta_0}/{\mathrm{d}\rho}$ and integrating over
all $\rho$, one can easily obtain the identity
\begin{equation} \label{eq:int-of-Wanis}
\int_0^\infty \sin^2\theta_0(\rho) \rho \mathrm{d}\rho =
\frac{2\omega_p}{\omega_0} \int_0^\infty
\left[1-\cos\theta_0(\rho)\right] \rho\mathrm{d}\rho,
\end{equation}
which gives a possibility to rewrite the soliton energy
\eqref{eq:Energy} as follows
\begin{equation}\label{eq:Energy-via-N}
\begin{split}
\mathscr{E} &= \mathscr{E}_{\text{exc}} + \hslash\omega_p N,\\
\mathscr{E}_{\text{exc}}  &= \frac{JS^2}{2}\int \mathrm{d}^2 x
\left[\left(\nabla\theta\right)^2 +
\sin^2\theta\left(\nabla\phi\right)^2\right].
\end{split}
\end{equation}
Note that the linear dependence of the soliton energy $\mathscr{E}$
on $N$ agrees with the general relation
$\hslash\omega_p=\delta\mathscr{E}/\delta N$.

The shape of the soliton essentially depends on the number $N$ of
bound magnons. In the case of solitons with large radius $R$
(equivalent to $N\gg N_0$), the approximate ``domain wall'' solution
works well, see Ref.~\cite{Kosevich90}. This solution has the shape
of a curved 1D domain wall at the distance $R$
\begin{equation}\label{eq:domain-wall}
\cos\theta_0(r) = \tanh\frac{r-R}{l_0}.
\end{equation}
Using this simple structure one can obtain the number of bound
magnons, which is proportional to the area of the soliton, $N\approx
N_0(R/l_0)^2$, and precession frequency
\begin{equation}\label{eq:Omega4large}
\frac{\omega_p}{\omega_0} \approx \frac{l_0}{R} \approx
\sqrt{\frac{N_0}{N}}.
\end{equation}
In the case of small radius solitons ($N\ll N_0$), the following
asymptotically exact solution works well \cite{Voronov83}
\begin{equation}\label{eq:smallR-structure}
\tan\frac{\theta_0(r)}{2} = \frac{R}{r_0}\ K_1
\left(\frac{r}{r_0}\right), \qquad r_0 =
\frac{l_0}{\sqrt{1-{\omega_p}/{\omega_0}}},
\end{equation}
where $K_1(\bullet)$ is the McDonald function. It provides correct
behavior for $r<R\ll l_0$, where it converts to the
Belavin--Polyakov solution $\tan \theta_0/2 = R/r$ and provides a
correct exponential decay for $r\gg R$. In this case the frequency
of the soliton precession $\omega_p\to\omega_0$ when $N\to0$, but
the dependence $\omega_p(N)$ has a singularity at the origin:
$d\omega_p/dN\to\infty$ as $N\to0$
\begin{equation}\label{eq:Omega4small}
\frac{\omega_p}{\omega_0} \approx 1-\frac{1}{\ln\left(8
N_0/e\gamma^2N\right)},
\end{equation}
where $\gamma\approx1.78$ is the Euler constant, see
Refs.~\cite{Voronov83,Ivanov86}.

In the intermediate case of arbitrary $R$, it is possible to use an
approximate trial function of the form, proposed in
Ref.~\cite{Sheka01},
\begin{equation}\label{eq:structure-trial}
\tan\frac{\theta_0(r)}{2} = \frac{R}{r} \exp
\left(-\frac{r-R}{r_0}\right).
\end{equation}
Here $R$ is the fitting parameter, which was found in
Ref.~\cite{Sheka01} by fitting the trial function
\eqref{eq:structure-trial} to the numerical solution of the
differential problem \eqref{eq:m-static}. The value of this fitting
parameter is closed to the soliton radius, which satisfies the
condition $\cos\theta_0(R)=0$.

The trial function \eqref{eq:structure-trial} gives a possibility to
describe approximately the soliton shape for a given radius $R$.
However, it contains one extra parameter, $\omega_p$, due to the
dependence $r_0 = r_0(\omega_p)$. One can calculate approximately
the $\omega_p(R)$--dependence as follows
\begin{equation} \label{eq:Omega(R)}
\omega_p(R) \approx \frac{\omega_0 l_0}{R+l_0},
\end{equation}
which provides the correct asymptote \eqref{eq:Omega4large} for
$R\gg l_0$ and gives the limiting value $\omega_p=0$ for $R\ll l_0$.
In the same approach the typical size of the exponential tail of the
soliton $r_0\approx l_0\sqrt{(R+l_0)/R}$; thus for the soliton shape
we have finally
\begin{equation}\label{eq:structure-trial'}
\begin{split}
\tan\frac{\theta_0(r)}{2} & \approx \frac{R}{r} \exp
\left(-\frac{r-R}{l_0}\sqrt{\frac{R}{R+l_0}}\right),\\
\phi & = \varphi_0 + \chi - \omega_p t.
\end{split}
\end{equation}

We will use this simple expression as initial condition for our
numerical simulations in Sec.~\ref{sec:simulations}.

\section{The Soliton Dynamics}
\label{sec:dynamics}

To describe the dynamics of the soliton as a whole,  it is necessary
first of all to introduce an effective soliton coordinate. Let us
define the soliton position $\vec{X}(t)= X(t) + iY(t)$ as the center
of mass of the z--component of the magnetization field:
\begin{equation} \label{eq:X-continuum}
\vec{X}(t) = \frac{S}{N a^2}\int \mathrm{d}^2x \ \vec{r}
\left(1-\cos\theta\right).
\end{equation}
Using this quantity we will look for the way  of possible soliton
deformation, which initialize its motion. In order to realize this
idea, let us derive the soliton speed (see the Appendix A):
\begin{equation} \label{eq:X'}
\frac{\mathrm{d}\!\vec{X}}{\mathrm{d}t} = \frac{JS^2}{\hslash N}\int
\mathrm{d}^2x \sin^2\theta\vec{\nabla}\phi .
\end{equation}
It is convenient to classify all possible perturbations of the
soliton shape using a complete set of functions. We choose the
solution of the linearized problem, which provides a set of partial
waves.

Let us remind that the soliton in an easy--axial ferromagnet has a
number of local magnon modes. The existence of local modes is
possible because of the gap in the magnon spectrum as predicted in
Ref.~\cite{Sheka01}; such modes correspond to different types of
soliton shape oscillations. To describe the local modes one has to
linearize the Landau--Lifshitz equations \eqref{eq:LL-continuum} on
the soliton background as it was done in Ref.~\cite{Sheka01}. We use
the partial--wave expansion
\begin{equation} \label{eq:modes}
\begin{split}
\theta  &= \theta_0(\rho) +\sum_m A_m
\left(u_m + v_m\right)\cos \Phi_m,\\
\phi  &= \varphi_0 + \chi - \omega_p t +\sum_m
\frac{A_m}{\sin\theta_0}\left(u_m - v_m\right) \sin \Phi_m,
\end{split}
\end{equation}
where $\Phi_m = m\chi-\omega_m t$ and $\omega_m$ is the magnon
frequency in the rotating frame. The radial functions $u_m(\rho)$
and $v_m(\rho)$ satisfy the following eigenvalue problem for two
coupled Schr\"odinger--like equations:
\begin{equation} \label{eq:EVP}
\begin{split}
\left[-\frac{\mathrm{d}^2}{\mathrm{d}\rho^2}-\frac{1}{\rho}
\frac{\mathrm{d}}{\mathrm{d}\rho} +
V_+(\rho)-\frac{\omega_m}{\omega_0}\right]u_m &= W(\rho) v_m,
\\
\left[-\frac{\mathrm{d}^2}{\mathrm{d}\rho^2}-\frac{1}{\rho}
\frac{\mathrm{d}}{\mathrm{d}\rho} +
V_-(\rho)+\frac{\omega_m}{\omega_0}\right]v_m &= W(\rho) u_m,
\end{split}
\end{equation}
where the radial ``potentials'' are
\begin{align*}
V_\pm(\rho) &= \frac{(\cos\theta_0 \pm m)^2}{\rho^2} +
\cos\theta_0\left(\cos\theta_0-\frac{\omega_p}{\omega_0}\right)\\
& - \frac12 \sin^2\theta_0\left(1+\frac{1}{\rho^2}\right) -
\frac12 (\theta^\prime_0)^2,\\
W(\rho) &= \frac12 \sin^2\theta_0\left(1+\frac{1}{\rho^2}\right) -
\frac12 (\theta^\prime_0)^2.
\end{align*}
The local modes exist in a range of frequencies inside the gap,
$\omega_m^{\text{loc}}\in(0,\omega_0 - \omega_p)$. The number of
local modes essentially depends on the soliton radius: when the
soliton radius decreases, the local modes leave the gap range,
transforming to the quasi--local modes with singularities in the
scattering picture. For the soliton with $R\leq R_c=1.52l_0$ there
exists only one local mode, namely the mode with $m=-1$, and it is
this mode which corresponds to the soliton motion.

Let us calculate the effective soliton trajectory  $\vec{X}(t)$
using the partial wave ansatz \eqref{eq:modes}. In the linear
approximation in $A_m$ all modes with $|m|\neq1$ give no
contribution to the integral \eqref{eq:X'} due to the angular
symmetry, and the effective soliton coordinat results as follows
(see the Appendix A for the details)
\begin{equation} \label{eq:X_orb}
\begin{split}
\vec{X}(t) &= R_{\text{orb}} e^{-i\omega_{-1}t},\; R_{\text{orb}} =
\frac{\left| A_{-1}\right|\ \left|C_{-1}\right| N_0 \omega_0 l_0}{2N
\omega_{-1}},\\
C_{-1}&=4\int_0^\infty \rho\mathrm{d}\rho \cos\theta_0 \Bigl(
u_{-1}v_1 - v_{-1}u_1\Bigr).
\end{split}
\end{equation}
Thus, only the perturbation with the symmetry of the mode $m=-1$ can
lead to a soliton motion as a whole. As we have found, the best way
to excite such a mode is to use the exact shape of this mode,
calculated in linear approximation, with \emph{finite} amplitude of
deformation $A_{-1}$. We will check this prediction in
Sec.~\ref{sec:dynamics-simulation}.

At the end of the section let us discuss the connection between the
dynamics of the soliton center $\vec{X}(t)$, which results in the
orbital motion \eqref{eq:X_orb}, and the dynamics of the specific
soliton position, introduced by \citet{Papanicolaou91} as a some
integral of the topological density $\mathcal{Q}$
\eqref{eq:Pontryagin}:
\begin{equation} \label{eq:R-Papanicolaou}
\vec{R} = \frac{\int \mathrm{d}^2x \vec{r} \mathcal{Q}}{\int
\mathrm{d}^2x \mathcal{Q}}.
\end{equation}
This quantity can be interpreted as a ``guiding center'' of the
soliton orbit. In the linear on $A_m$ approximation one can
calculate (see the Appendix A) that
\begin{equation} \label{eq:R-result}
\vec{R}(t)= - A_1l_0\vec{e}_{x} = \vec{\mathrm{const}},
\end{equation}
which corresponds to the simple soliton shift due to the
translational mode with $m=+1$. The physical picture is similar to
the electron motion in the magnetic field: the electron moves along
the circular Larmor orbit and $\mathrm{d}\vec{X}/\mathrm{d}t$ is not
conserved; the generalized momentum $\vec{P}$ also changes; however
their combination, which determine the center of the ``guiding
center'' of the orbit $\vec{R}$ saves its position.

\section{Numerical Simulations for the Circular Symmetric
Topological Precessional Soliton} \label{sec:simulations}

To validate predictions of the continuum theory for the soliton
properties, we integrate numerically the discrete Landau--Lifshitz
equations \eqref{eq:LL-discrete} over square lattices of size
$L\times L$ using a 4th--order Runge--Kutta scheme with time step
$0.01$ and periodic boundary conditions. In all cases the soliton is
started near the center of the domain. We have fixed the exchange
constant $J=\hslash=1$ as well as the spin length $S=1$. We have
considered the anisotropy parameter in the range
$\delta\in(0.0005;0.1)$, corresponding to $l_0/a \in (22.4,1.58)$ so
that we are close to the continuum limit.  We consider system sizes
in the range $L/a\in(50,800)$.

\begin{figure}
\begin{center}
\includegraphics[width=\columnwidth]{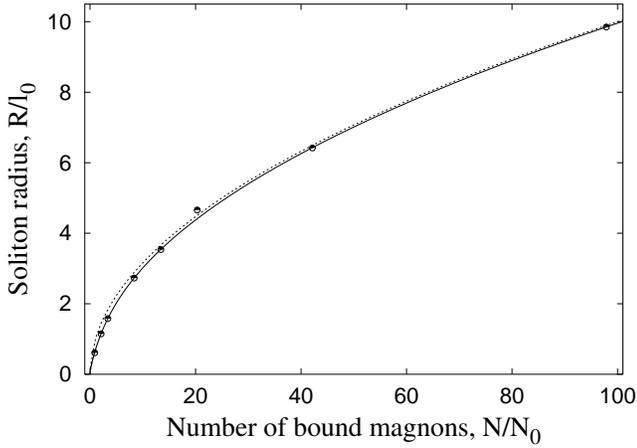}
\end{center}
\caption{Radius of the soliton as a function of the number of bound
magnons. The symbols correspond to the simulation data; the solid
line is the result of the numerical integration of the continuum
Eqs.~\eqref{eq:m-static}, \eqref{eq:N'}; the dashed line corresponds
to Eq.~\eqref{eq:N=R^2}} \label{fig:R_via_N}
\end{figure}

We start the simulations using an initial soliton--like distribution
\begin{equation} \label{eq:soliton-ini}
\theta = \theta_0(r), \quad \phi = \varphi_0 + \chi,
\end{equation}
with the trial function \eqref{eq:structure-trial'} for the
$\theta$--field. Evidently, the soliton solutions of the
Landau--Lifshitz equations for a lattice differ from the circular
symmetric continuous solutions, and also from the simple trial
function. To find such a ``pure'' soliton solution, i.e. to adapt
the trial solution to the lattice, one should provide enough time
for the decay of the initial error in the trial functions. In fact,
using \eqref{eq:soliton-ini} as initial conditions for the lattice
we excite also magnons, which should be taken out the system. To
avoid the problem of magnons we have damped them the initial stage
of simulations by applying damping. This kills all spreading spin
waves coming from the imperfect initial condition. In this way
instead of Eqs.~\eqref{eq:LL-discrete}, we have integrated
numerically Landau--Lifshitz equations with Gilbert damping
\begin{equation*} %\label{eq:LL-discrete(num)}
\hslash (1 + \varepsilon^2) \frac{d \vec{S}_{\vec{n}} }{dt}= -
\left[\vec{S}_{\vec{n}} \times \frac{\partial \mathscr{H} }{\partial
\vec{S}_{\vec{n}}}\right] + \frac{\varepsilon}{S} \left[
\vec{S}_{\vec{n}}\times \left[\vec{S}_{\vec{n}} \times
\frac{\partial \mathscr{H} }{\partial
\vec{S}_{\vec{n}}}\right]\right],
\end{equation*}
see Ref.~\cite{Sheka05a} for details. The lowest frequency of the
continuous magnon spectrum is $\omega_0$, thus the damping time
$t_d\approx 1/(\varepsilon \omega_0)$, see Ref.~\cite{Kamppeter01}
for details. During the damping time ($t<t_d$), the magnons are
damped in the system, but the soliton is also damped, and the
soliton energy $\mathscr{E}$ decays as well as the number of bound
magnons $N$. In order to save the soliton structure, we should
switch off the damping before we damp out the soliton, i.e. $t <
1/\varepsilon \omega_p$. In all simulations we use the same value of
$\varepsilon=0.02$, then the damping time $t_d\approx 12/\delta$,
and the damping is turned off \emph{adiabatically} after a time
greater than $t_d$.

Let us discuss the choice of the other parameters. In all
simulations we want to be not far from the continuum limit in order
to validate the continuum approach. It means that the magnetic
length $l_0$ should be greater than the lattice constant $a$. This
regulates the choice of the anisotropy constant $\delta =
a^2/4l_0^2$. Besides $l_0$ the soliton shape is characterized by two
extra scales: $R$, which is the soliton radius, and $r_0$, which
characterizes the scale of the exponential decay of the excitation
far from the soliton center.

We start with the large radius solitons. In this case $R\gg r_0
\approx l_0$, and we can limit ourselves by choosing $\delta=0.1$
(this corresponds to $l_0\approx 1.6 a$). The system size $L$ should
be much greater than the largest parameter of the soliton, which is
its radius. We consider solitons up to the radius $R=20l_0\approx
31.6 a$. Thus we consider lattices with $L=200a$ , which satisfy all
above mentioned  conditions.

\begin{figure}
\begin{center}
\includegraphics[width=\columnwidth]{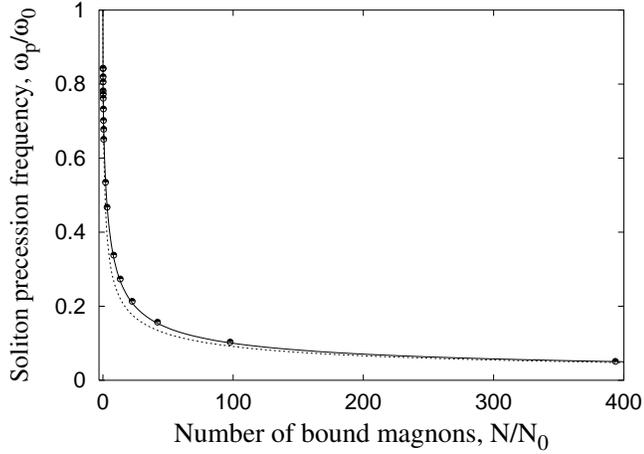}
\end{center}
\caption{Soliton precession frequency as a function of the number of
bound magnons. Symbols correspond to the simulation data; the solid
line is the result of continuum model integration; the dashed line
corresponds to Eq.~\eqref{eq:Omega(N)}.} \label{fig:Omega_via_N}
\end{figure}

In the case of small radius solitons we have the following relation
between the parameters of the system:
\begin{equation} \label{eq:small-uneqal}
a \ll R \ll l_0 \ll r_0 \ll L.
\end{equation}
For the smallest soliton we choose $\delta=0.0005$, which
corresponds to $l_0\approx 22.4a$; this gives the possibility to
consider solitons of small radii down to $R=0.225l_0\approx 5a$.
However, such a small anisotropy drastically changes the soliton
shape far from the center, which has the scale $r_0$, see
Eq.~\eqref{eq:smallR-structure}. For example, for the soliton with
$R=0.225 l_0$, the precession frequency $\omega_p\approx
0.84\omega_0$ (see Ref.~\cite{Sheka01}), which results in
$r_0=l_0/\sqrt{1-\omega_p/\omega_0}\approx56a$. Thus to consider
small radius solitons we must increase the system size. In our spin
dynamics simulations we choose $L=800a$  for the smallest solitons.
To perform simulations for such large systems, $800\times 800$, we
have used parallelize a computations, see the Appendix B.

\begin{figure}
\begin{center}
(a) \hspace{0.4\columnwidth} (b) \\
\includegraphics[width=0.47\columnwidth]{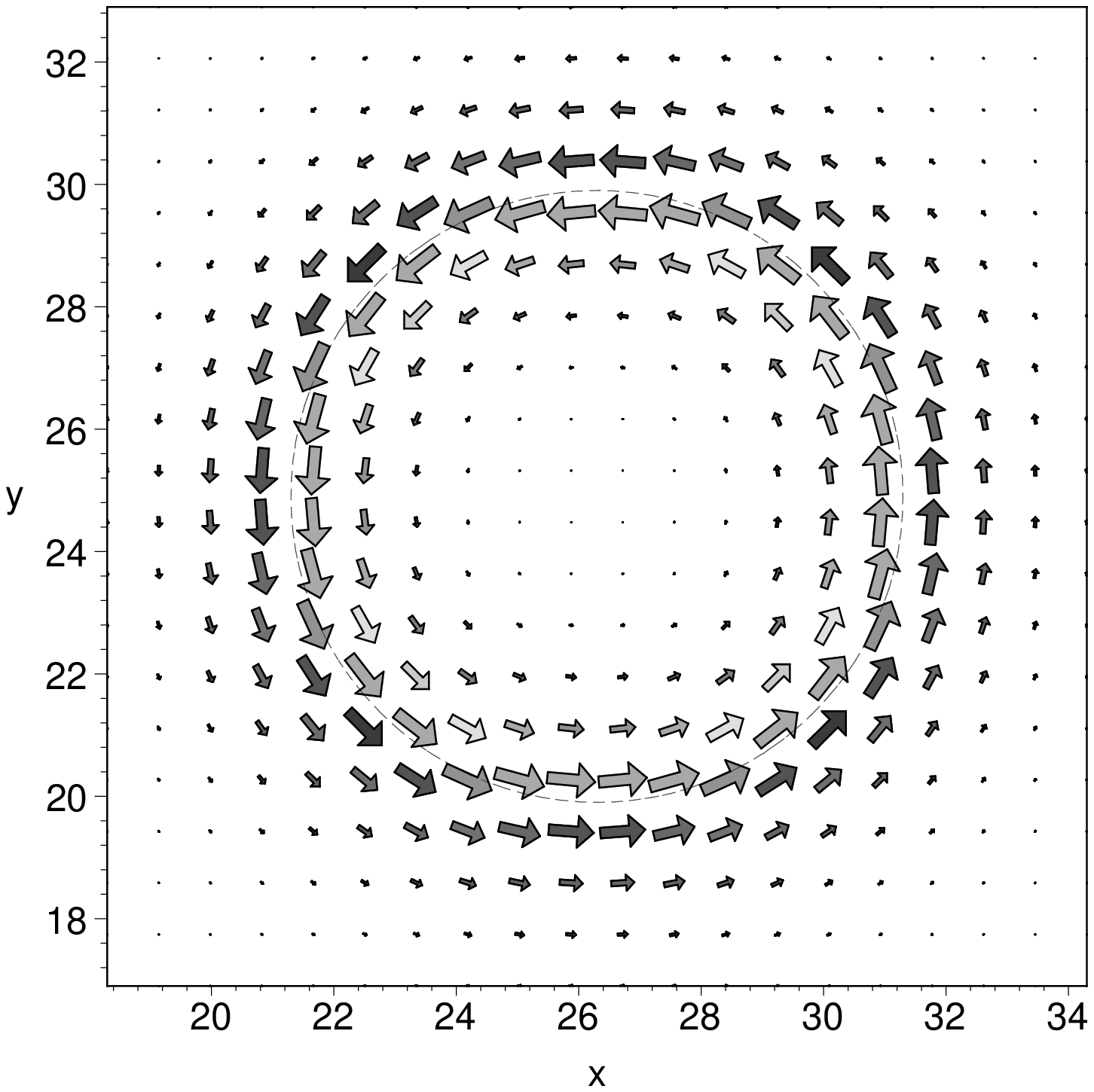}
\includegraphics[width=0.47\columnwidth]{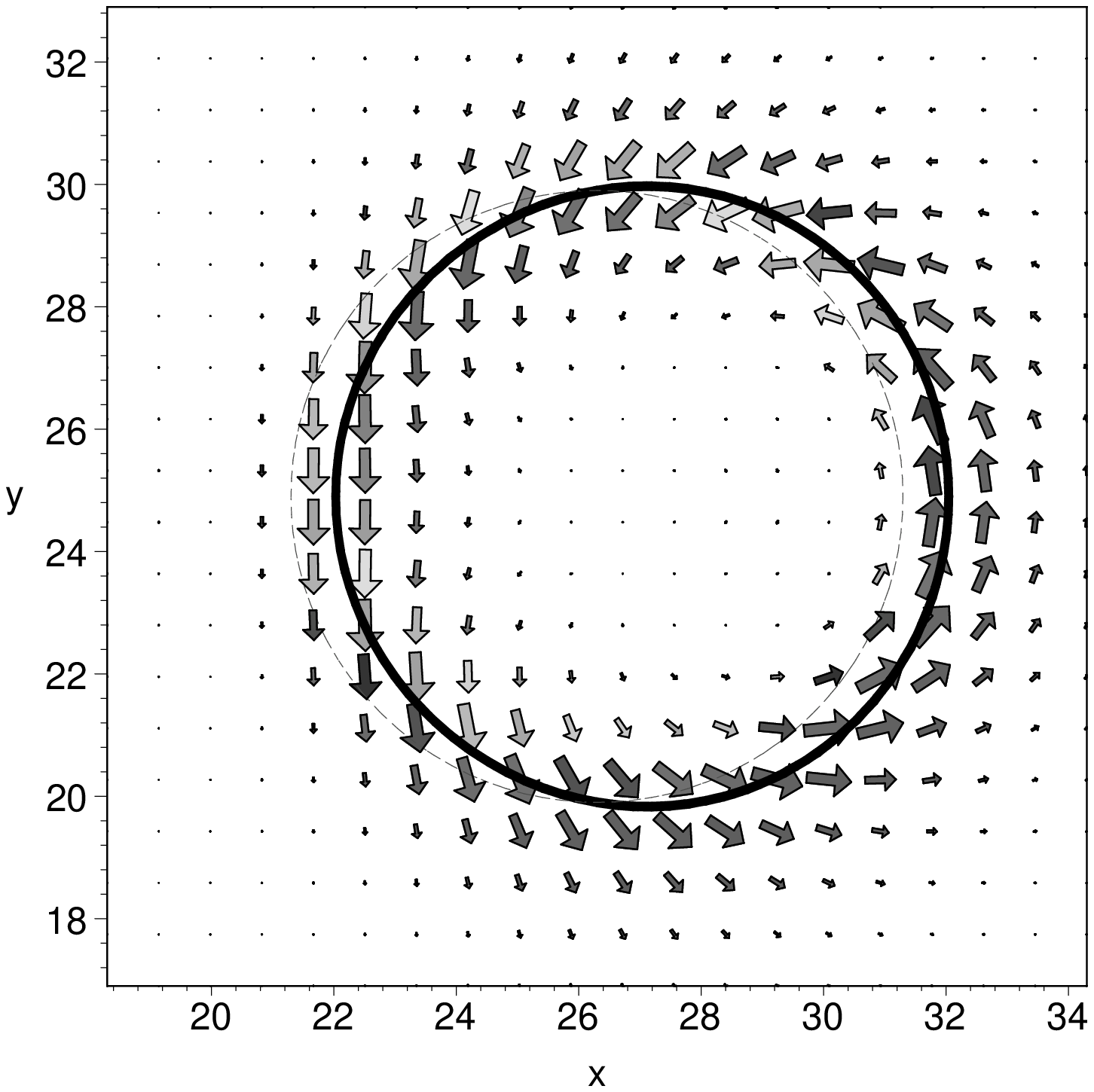}
\end{center}
\caption{In--plane spin distribution for the soliton with $R=5l_0$,
which is situated at $\vec{X}_0 = 26.3 + i 25.0$. Fig.(a)
corresponds to the circular symmetric soliton and (b) to an
elliptically deformed one. The lines describe the contour plot of
$S_z=0$ ($\theta=\pi/2$): the dashed line is for the circular
symmetric soliton and the solid line is for the deformed soliton.}
\label{fig:in_plane}
\end{figure}

Let us discuss results of our spin dynamics simulations. Starting
from initial conditions \eqref{eq:soliton-ini}, and adapting the
soliton shape to the lattice, we have obtained the class of
one--parameter stable soliton solutions for a wide range of the
parameter $N$, or equivalently, the soliton radius $R$. We have
studied the $R(N)$ dependence, which is presented in Fig.
\ref{fig:R_via_N}. Almost in the full range of parameters, the
simple dependence
\begin{equation} \label{eq:N=R^2}
N\approx N_0 \left(\frac{R}{l_0}\right)^2
\end{equation}
is valid. Note that the dependence \eqref{eq:N=R^2} was verified
numerically in Ref.~\cite{Kamppeter01} for the large radius solitons
only, where $R>10l_0$. Here we want to check the continuum results
for arbitrary $R$. Using \eqref{eq:Omega(R)} and \eqref{eq:N=R^2},
an approximate dependence
\begin{equation} \label{eq:Omega(N)}
\omega_p(N) \approx \frac{\omega_0}{1+\sqrt{N/N_0}}
\end{equation}
can be derived. To compute the precession frequency, we calculate
the Fourier spectrum of the in--plane spin components. One can see
from Fig.~\ref{fig:Omega_via_N} that this simple dependence works
well in a wide range of parameters.

\section{Simulation of the Orbital Motion of the Soliton}
\label{sec:dynamics-simulation}

\begin{figure}
\begin{center}
\includegraphics[width=\columnwidth]{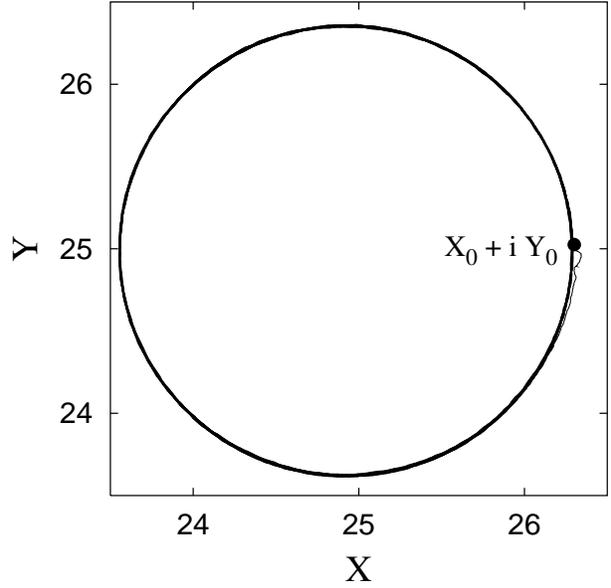}
\end{center}
\caption{Trajectory of the soliton with initial radius $R=4.87 l_0$
and initial precession frequency $\omega_p=0.2\omega_0$. The initial
point $X_0 + iY_0$ corresponds to the center of the soliton from the
Fig.~\ref{fig:in_plane}.} \label{fig:orbit}
\end{figure}

In the previous section we have performed spin dynamics simulations
only for the circular symmetric precessional soliton, which does not
move as a whole. As mentioned in Sec.~\ref{sec:dynamics}, in order
to move the soliton one should break its symmetry. We have done this
by an initial deformation of the soliton shape, and integrated
numerically the Landau--Lifshitz equations. Specifically, we have
chosen an elliptic kind of deformation, which corresponds to the
shape of the internal partial mode with azimuthal quantum number
$m=-1$. We start the simulations with initial conditions
\begin{equation} \label{eq:m=-1-mode} %
\begin{split}
\theta  &= \theta_0(\rho) + A [u_{-1}(\rho)+v_{-1}(\rho)]\cos\chi,\\
\phi  &= \varphi_0 + \chi - \frac{A}{\sin\theta_0}
[u_{-1}(\rho)-v_{-1}(\rho)]\sin\chi
\end{split}
\end{equation}
by the same numerical scheme as described in the previous section.
We calculated the functions $u_{-1}(\rho)$ and $v_{-1}(\rho)$
numerically solving the eigenvalue problem \eqref{eq:EVP} by the
two--parametric shooting scheme as described in Ref.~\cite{Sheka01}.
The parameter $A$ is the amplitude of the eigenmode, which
characterizes the magnitude of the soliton deformation. An initial
distribution of spins, which corresponds to
Eqs.~\eqref{eq:m=-1-mode}, is shown in Fig.~\ref{fig:in_plane} and
can be seen to describe the elliptical kind of the soliton
deformation.

During the simulations we have computed the time dependence  of the
position $\vec{X}(t)=\bigl(X(t),Y(t)\bigr)$ of the soliton center:
\begin{equation} \label{eq:X-discrete}
\vec{X}(t) = \frac{\sum_{\vec{n}}\vec{r}_{\vec{n}}
\left[S-S^z_{\vec{n}}(t) \right]}{\sum_{\vec{n}}S-S_{\vec{n}}^z(t)},
\end{equation}
which is the discrete analogue of Eq.~\eqref{eq:X-continuum};
$\vec{r}_{\vec{n}}=(x_n, y_n)$ are the lattice points.

\begin{figure}
\begin{center}
\includegraphics[width=\columnwidth]{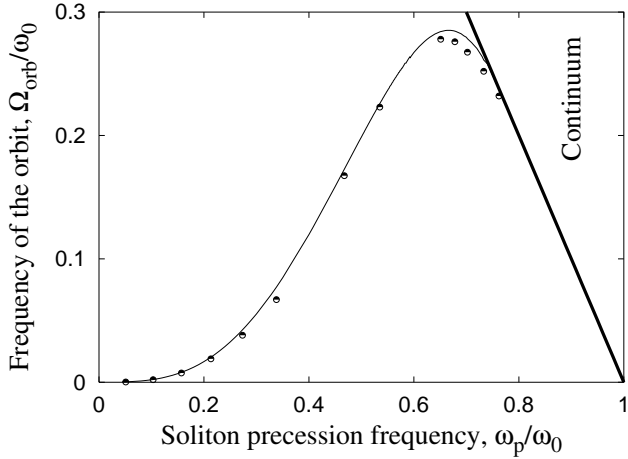}
\end{center}
\caption{Frequency of the orbit motion as a function of the
precession frequency of the soliton. Symbols correspond to the spin
dynamics simulation data; lines corresponds to the frequency of the
eigenmode with $m=-1$ from the continuum theory by \citet{Sheka01}.
\label{fig:omega_via_Omega}}
\end{figure}

\begin{figure}
\begin{center}
\includegraphics[width=\columnwidth]{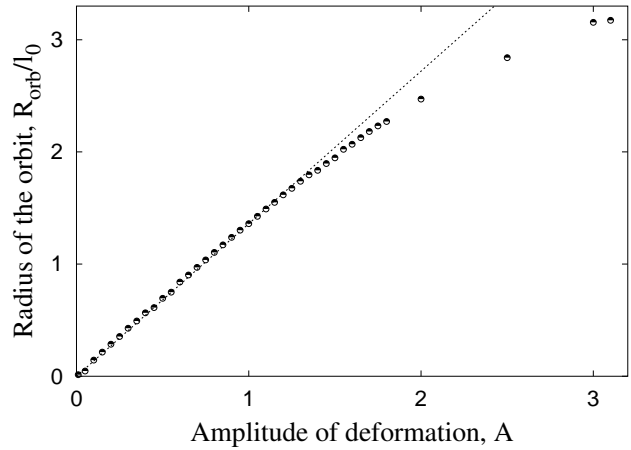}
\end{center}
\caption{Radius of the soliton orbit as a function of the
deformation amplitude $A$. Parameters of the soliton:  initial
radius $R=4.87 l_0$ and initial precession frequency
$\omega_p=0.2\omega_0$. \label{fig:R-via-A}}
\end{figure}

We have found numerically that after switching off the damping, the
soliton reaches very fast a circular trajectory, see
Fig.~\ref{fig:orbit}. This results in a nice circular motion with
constant frequency. For small deformations the radius of the orbit
is proportional to the initial deformation. One can say that the
effect of a circular motion and the excitation of the mode with
$m=-1$ are identical for this case, as predicted by \citet{Sheka01}.
Such a relation is valid in some range of the soliton deformation
for all soliton radii, see Fig.~\ref{fig:omega_via_Omega}. Then, for
larger deformations, non--linear regime is clearly seen, see
Fig.~\ref{fig:R-via-A}. The frequency of this orbit motion of the
soliton approximately corresponds to the frequency of the local
mode, $\Omega_{\text{orb}} = \omega_{m=-1}.$

The presence of such an exact circular motion, with only one
frequency, independent of the orbit radius (even in non--linear
regime), gives the possibility to conclude that this is the
\emph{first} observation in the numerical experiment of the pure
gyroscopic motion, which is equivalent to the Larmor precession of a
charged particle in a magnetic field. Therefore, the soliton motion
can be described by an effective equation of motion for the position
of the soliton $\vec{X}$, which takes the form of usual Newtonian
equation for a particle with the \emph{well--defined effective mass}
$M$ under the influence of the gyroscopic force $\vec{F}_g$:
\begin{equation} \label{eq:EOM}
M\frac{\mathrm{d}^2\!\vec{X}}{\mathrm{d}t^2} =  \vec{F}_g, \qquad
\vec{F}_g = G\left[\vec{e}_z \times
\frac{\mathrm{d}\!\vec{X}}{\mathrm{d}t} \right].
\end{equation}
Here $G=4\pi\hslash S/a^2$ is the gyroconstant,  see
\cite{Kosevich90}. Formally, Eq.~\eqref {eq:EOM} has two solutions.
One solution, $\vec{X}=\text{const}$, corresponds to the translation
mode with $m=+1$. In the infinite system this is a zero--frequency
local mode, which describes a simple shift of the soliton,
$\omega_{m=+1}=0$. The second solution describes a circular motion
with the frequency $\Omega_{\text{orb}} = G/M$. Thus, we can
calculate the effective mass of the soliton using the simulation
data for the orbit frequency.

We have checked how the orbit frequency depends on the soliton
radius. For the large radius solitons $\Omega_{\text{orb}} \approx
2\omega_0(l_0/R)^3$ is in good agreement with the results for the
local modes. For the small radius solitons, the orbit frequency
tends to the boundary of the spectrum, $\Omega_{\text{orb}}\approx
\omega_0-\omega_p$. This dependence corresponds to our result for
the eigenfrequencies, see Eq.~(16) of the Ref.~\cite{Sheka01}. In
the case $N\ll N_0$ one can use the approximate limiting expression
\eqref{eq:Omega4small} for $\omega_p$, which results in
$\Omega_{\text{orb}} \approx {\omega_0}/{\ln\left(8
N_0/e\gamma^2N\right)}$. The frequency tends to zero when $R\to0$.

One can calculate the effective mass of the soliton by  $M =
{G}/{\Omega_{\text{orb}}} $,
\begin{equation}\label{eq:mass}
M = M_0 F\left(\frac{N}{N_0}\right), \qquad M_0 = \frac{G}{\omega_0}
= \frac{\pi\hslash^2}{Ja^2\delta},
\end{equation}
where $M_0$ is a characteristic value of the effective mass, found
in Ref. ~\cite{Ivanov89}. The function $F(\bullet) =
\omega_0/\Omega_{\text{orb}}(\bullet)$ depends on the soliton size
having the asymptotic behavior
\begin{equation}\label{eq:F(x)}
F(x) =
  \begin{cases}
    \ln\left(\dfrac{8}{e\gamma x}\right) & \text{when $x\ll 1$}, \\
    \frac12 x^{3/2} & \text{when $x\gg1$}.
  \end{cases}
\end{equation}
The soliton mass diverges in the limiting cases when $R\to0$ and
$R\to\infty$. Note that for the large radius solitons the mass
increases faster that the domain wall width, which is proportional
to $R$. The mass takes a minimum value $M_c\approx 3.51 M_0$ for the
soliton with $R_c\approx 0.547 l_0$. The soliton with these
parameters has the highest \emph{mobility}.

\section{Conclusion}
\label{sec:conclusion}

In this paper we have studied the dynamics of topological solitons
in classical 2D easy--axis ferromagnet. The analysis was made both
analytically in the continuum approximation and numerically using
the spin dynamics simulations for a wide range of solitons: from
large radius solitons to small ones. Our simulations were performed
for small anisotropies, which corresponds to the continuum
description. We have checked and confirmed a number of results from
the continuum theory about the soliton structure, in particular, the
connection between the number of bound magnons and the precession
frequency of the spins inside the soliton.

The main issue is connected  with the soliton dynamics. We have
proposed a way how to move a soliton exciting one of its internal
modes. To our knowledge, it is the first observation of inertial
motion of 2D magnetic solitons. The effective soliton dynamics is
similar to the Larmor dynamics of a charged particle in a magnetic
field. By analysis of the effective soliton dynamics we extract
information about the effective mass of the soliton. This mass
essentially depends on the anisotropy, $M\propto 1/\delta$, and on
the soliton size, having the minimum for the soliton of the radius
about $0.5 l_0$. In the case of large radius solitons the soliton
mass increases with the increase of the soliton radius. Note that it
increases faster than the number of the bound magnons, $M\approx
\frac12M_0(N/N_0)^{3/2}$. Such dependence is in a good agreements
with previous results \cite{Ivanov89}; the soliton mass diverges in
the limit case $R\to \infty$, which corresponds to the fact that the
single domain wall can not move. When the soliton radius is smaller,
the effective mass increases logarithmically with its radius, and
diverges in the limit case $R=0$. Note that the problem of inertial
properties of a small radius soliton have caused a lot of
discussions. According to linear analysis \cite{Ivanov89}, the
soliton mass tends to some limit value
$M^\star=6\sqrt{\pi}(\pi+2)M_0\approx55M_0$ when $R\to0$. At the
same time our previous analysis of eigenmodes \cite{Sheka01}  shows
that $M\to\infty$ at the limit case $R\to0$. Spin dynamics
simulations confirms our results on internal modes: the soliton
loose its mobility when becomes very small.

We have predicted the fine circular motion of the  soliton by
exciting its internal mode. We believe that such phenomenon can be
observed experimentally, e.g. by ac pumping. Our investigations can
be important also for the quantum Hall systems, where skyrmion--type
solitons are well-known to lead to the breakdown of the
spin--polarized quantum Hall effect \cite{Cooper98}.

\begin{acknowledgement}

D.~D.~Sheka thanks the University of Bayreuth, where part of this
work was performed, for kind hospitality and acknowledges support
from the Alexandr von Humboldt Foundation.
%Deutsches Zentrum f{\"u}r Luft- und Raumfart e.V., Internationales
%B{\"u}ro des Bundesministeriums f{\"u}r Forschung und Technologie,
%Bonn, in the frame of a bilateral scientific cooperation between
%Ukraine and Germany, project No. UKR--02--011. D.~D.~Sheka
%acknowledge also the support from Ukrainian--French ``Dnipro'' grant
%(No. 09855WF).
C.~Schuster thanks Stefan Karpitschka for his kind support
concerning aspects of parallelizing the source code.

\end{acknowledgement}

\appendix

\section*{Appendix A\ Calculation of the soliton velocity}
\label{sec:dX/dt}

In order to check the possible soliton motion we calculate the speed
of the effective soliton position $\vec{X}(t)$. According to the
Eq.~\eqref{eq:X-continuum} the soliton speed
\begin{align*}
\frac{\mathrm{d}X_i}{\mathrm{d}t}& = \frac{S}{N a^2}\int
\mathrm{d}^2 x\ x_i \sin \theta \partial_t \theta = \frac{1}{\hslash
N}\int \mathrm{d}^2 x\ x_i \frac{\delta \mathscr{E}}{\delta\phi},
\end{align*}
where we used the Landau--Lifshitz Eq.~\eqref{eq:LL-continuum}.
Calculating the functional derivative for the energy functional
\eqref{eq:Energy} and integrating by parts using the identity
\begin{equation*}
x_i \vec{\nabla}\cdot\left(\sin^2\theta\vec{\nabla}\phi\right) =
\vec{\nabla}\cdot\left(x_i\sin^2\theta\vec{\nabla}\phi\right) -
\sin^2\theta\partial_i \phi,
\end{equation*}
one can derive the soliton velocity in the form
\begin{equation} \label{eq:X'(1)}
\frac{\mathrm{d}\!\vec{X}}{\mathrm{d}t} = \frac{2\pi l_0
JS^2}{\hslash N}\int_0^\infty\!\! \rho\mathrm{d}\rho \Bigl\langle
\sin^2\theta\vec{\nabla_{\!\!\rho}} \phi \Bigr\rangle,
\end{equation}
which is equivalent to Eq.~\eqref{eq:X'}. Here the averaging means
$\bigl\langle F(\bullet,\chi)\bigr\rangle \equiv
(1/2\pi)\int_0^{2\pi}F(\bullet,\chi)\mathrm{d}\chi$.

Using the partial--wave ansatz \eqref{eq:modes}, one can concretize
the average value in Eq.~\eqref{eq:X'(1)}:
\begin{align*}
&\Bigl\langle \sin^2\theta\vec{\nabla_{\!\!\rho}} \phi \Bigr\rangle
= \sum_m A_m\! \Biggl\{ \!\sin^2\theta_0 \left[\frac{u_m -
v_m}{\sin\theta_0}\right]^\prime \Bigl\langle
\sin \Phi_m e^{i\chi} \Bigr\rangle\\
&+\frac{i}{\rho}\left[ (u_m+v_m) \sin2\theta_0 + m (u_m-v_m)\sin
\theta_0 \right]\Bigl\langle \cos \Phi_m e^{i\chi}
\Bigr\rangle\Biggr\} .
\end{align*}
After averaging with account of the expressions
\begin{equation} \label{eq:average}
\begin{split}
\Bigl\langle \cos \Phi_m e^{i\chi} \Bigr\rangle &=
\frac{\delta_{|m|,1}}{2}e^{im\omega_m t},\\ \Bigl\langle \sin \Phi_m
e^{i\chi} \Bigr\rangle &= \frac{im\delta_{|m|,1}}{2}e^{im\omega_m
t},
\end{split}
\end{equation}
one can calculate the soliton velocity in the form
\begin{equation} \label{eq:X'-result-1}
\frac{\mathrm{d}\!\vec{X}}{\mathrm{d}t} = \frac{i\omega_0  l_0
N_0}{2N}\Bigl(A_1 C_1 e^{i\omega_{1}t} + A_{-1}C_{-1}
e^{-i\omega_{-1}t}\Bigr),
\end{equation}
where the constants $C_m$ are determined by the static soliton
structure,
\begin{equation*}
\begin{split}
C_m = -2\!\int_0^\infty \!\! \rho\mathrm{d}\rho \cos\theta_0
\Biggl\{&(1+m)\Bigl(u_mu_1 - v_mv_1\Bigr)\\
& - (1-m)\Bigl(u_mv_1 - v_mu_1\Bigr)\!\Biggr\}\!.
\end{split}
\end{equation*}
The mode with $m=1$ is the zero--frequency local mode,  which
describes a shift of the soliton position, its eigenspectrum has the
form \cite{Sheka01}
\begin{equation*}
u_1 = \theta_0^\prime -\frac{\sin\theta_0}{\rho},\quad v_1  =
\theta_0^\prime +\frac{\sin\theta_0}{\rho}, \qquad \omega_1 = 0.
\end{equation*}
A simple calculation shows that $C_1=0$, therefore the soliton
motion is connected only with the mode $m=-1$, see
Eq.~\eqref{eq:X_orb}.

Let us calculate the dynamics of the ``guiding center'' position of
the soliton \eqref{eq:R-Papanicolaou}. Using the partial--wave
expansion \eqref{eq:modes}, one can rewrite the topological density
$\mathcal{Q}$ \eqref{eq:Pontryagin} as follows:
\begin{align*}
\mathcal{Q} = &-\frac{\sin\theta_0\, \theta_0^\prime}{l_0^2\rho}  -
\frac1{l_0^2\rho} \sum_m A_m \cos\Phi_m
\Bigl[ (u_m^\prime+v_m^\prime) \sin\theta_0\\
& + (u_m + v_m) \cos\theta_0 \theta_0^\prime + m (u_m - v_m)
\theta_0^\prime \Bigr].
\end{align*}
Averaging the linear momentum \eqref{eq:R-Papanicolaou} with account
of \eqref{eq:average} one can derive $\vec{R}(t)$ in the form
\begin{equation} \label{eq:R}
\vec{R}(t) = l_0A_1C_1^\star + l_0A_1C_{-1}^\star \exp
\left(-i\omega_{-1}t\right),
\end{equation}
where the constant $C_m^\star$ can be calculated as follows:
\begin{equation*}
\begin{split}
C_m^\star = \frac14\!\int_0^\infty \!\! \rho\mathrm{d}\rho
\Biggl\{&(1+m)\Bigl(u_mu_1 - v_mv_1\Bigr)\\
& - (1-m)\Bigl(u_mv_1 - v_mu_1\Bigr)\!\Biggr\}\!.
\end{split}
\end{equation*}
One can easily see that $C_1^\star=-1$, and the contribution of the
mode $m=+1$ results in the soliton shift. For the mode with $m=-1$
one can rewrite the constant $C_{-1}^\star$ in the form:
\begin{equation*}
C_{-1}^\star = \frac12\!\int_0^\infty \!\! \rho\mathrm{d}\rho
\Bigl(u_{-1}v_1 - v_{-1}u_1\Bigr).
\end{equation*}
This integral vanishes due to the symmetry of the eigenvalue problem
\eqref{eq:EVP}. Namely, a simple calculation shows that
\begin{align*}
&\frac{\omega_{m} + \omega_{-m}}{\omega_0}\Bigl(u_mv_{-m} - v_m
u_{-m}\Bigr)\\
&= \vec{\nabla}\cdot\Bigl(u_m\vec{\nabla}v_{-m} +
v_m\vec{\nabla}u_{-m} - u_{-m}\vec{\nabla}v_{m} -
v_{-m}\vec{\nabla}u_{m} \Bigr)
\end{align*}
for any $m$. The righthandside is in the form of the total
divergency, thus it gives no contribution to the integral over the
system:
\begin{equation*}
\int_0^\infty \!\! \rho\mathrm{d}\rho \left(u_mv_{-m} - v_m
u_{-m}\right) = 0.
\end{equation*}
Therefore the constant $C_{-1}^\star=0$, and finally the ``guiding
center'' of the soliton can be rewriting in the form
\eqref{eq:R-result}.

\section*{Appendix B\ Parallelized Computations}
\label{sec:parallel}

There exist two main possibilities of parallelization computations:
(i) the usage of a vector--computing machine which has a shared
memory or (ii) the usage of a cluster system (usually a Linux
cluster) with communication between the different processors
(nodes).  The latter process is called \emph{message passing
interface} (MPI), this interface does not depend on the programming
language.

In MPI there is a master--process which  is responsible for the
administration of the data, i.e. initializations, reading or saving
of data, whereas the other nodes (slaves) are doing the calculation
(e.g. integration). In our case we divide the lattice into
horizontal stripes. As in our system we take into account
nearest-neighbor interaction we must put a communication between the
borders of the stripes. To make calculations for the $i-th$ stripe
we need the lower border of the $(i+1)-$th stripe and the upper
boarder of the $(i-1)-$th stripe. This exchange is done after every
second integration step in the Runge Kutta algorithm. This
advantages a good equilibrium between the latency period (time in
which the nodes are synchronized), data-transferring time, and
calculation time. As the boundary condition of our lattice is
periodic the upper boarder of the top stripe is exchanged with the
lower border of the bottom stripe. Concerning the left and right
boarders of one stripe there are no communication processes, because
horizontal cuts of the system advantage an internal exchange of the
lateral borders in one process.

As the data is always transferred between the same nodes we
implement a \emph{persistant connection mode} in order to make the
\emph{overhead} (additional time spent on the connection
establishment) smaller. Theses connections are built up at the
beginning of the integration to be ready for a fast use.

The transfer takes place in an asynchronous,  buffered mode. Thereby
the data resulting of the calculation on a stripe is put into a
buffer so that for this node there is no need to wait until the
other nodes are ready to receive. In order to do a new calculation
in the next time steps the stripe awaits information of its
neighbored stripes. The advantage of this method is that the time
exposure for the synchronization of the nodes decreases.

After integrating 40 time steps with  $\Delta t=0.01$ we calculate
the soliton position $\vec{X}(t)$ discretely and optimally
distributed over the other nodes. After finishing the calculations
the data is sent from the slaves to the master-process which saves
the data in a file.

%The simulation is interrupted as soon as the master-process finds in
%the working directory a file which we  arbitrarily call
%''VMDCANCEL''. But before the simulation stops all spin data are
%sent to the master-process which writes the complete spin-field into
%a file. This advantages the simulation to continue from this last
%time step anytime later.

The parallelizing of the source code  is much easier and more
effective if one can use shared memory multi-processors machines.
The loops are environed by sunstyle parallelization-directives.
According to the dependency of the variables they have to be
declared as \emph{private variables} and others as \emph{reduction
variables}. \emph{Private variables} take different values in
different threads (such as auxiliary variables) so there is no
shared memory for these variables. \emph{Reduction variables} are
used for sums which are distributed over different threads during
one loop and are added in the end of the parallelized loop. The
advantage of this method is a very fast access to the shared memory
of all threads. This avoids a time consuming latency period caused
by a communication process. To summarize we want to stress that a
shared memory machine is always much faster than a cluster, if one
considers a fixed number of processors to be in use. The only
advantage a cluster has is the fact that compared to a shared memory
machine it has much more processors and the number of processors can
be increased arbitrarily. Also aspects of the price motivate
universities and institutes to buy a cluster instead of a shared
memory machine.

%\bibliography{soliton}
\end{document}